\documentclass[twocolumn,aps,prd,amssymb,eqsecnum,nofootinbib,superscriptaddress,floatfix]{revtex4}
\newlength{\picwidth}
\usepackage{epsfig}
\usepackage{amsmath}
\usepackage{dcolumn}
\usepackage{amssymb}
\usepackage{amsfonts}
\usepackage{enumerate}
\usepackage{bm}
\begin{document}
\def\be{\begin{equation}}
\def\ee{\end{equation}}
\def\ba{\begin{eqnarray}}
\def\ea{\end{eqnarray}}

\title{Lema\^{i}tre-Tolman-Bondi cosmological models, smoothness, and positivity
of the central deceleration parameter}

\author{R. Ali Vanderveld}
\affiliation{Jet Propulsion Laboratory, California Institute of Technology, Pasadena, CA 91109}
\author{\'{E}anna \'{E}. Flanagan}
\affiliation{Center for Radiophysics and Space Research, Cornell University, Ithaca, NY 14853}
\affiliation{Laboratory for Elementary Particle Physics, Cornell University, Ithaca, NY 14853}
\author{Ira Wasserman}
\affiliation{Center for Radiophysics and Space Research, Cornell University, Ithaca, NY 14853}
\affiliation{Laboratory for Elementary Particle Physics, Cornell University, Ithaca, NY 14853}
\date{\today}
\begin{abstract}

We argued in a previous paper [R.\ A.\ Vanderveld {\it et al.}\ 2006] that
negative deceleration parameters at the center of symmetry in
Lema\^{i}tre-Tolman-Bondi cosmological models can only occur if the model
is not smooth at the origin.  Here we demonstrate explicitly the
connection between non-smoothness and the failure of positivity
theorems for deceleration.  We also address some confusion that has
arisen in the literature and respond to some recent criticisms of our
arguments.

\end{abstract}
%\pacs{ 98.80.-k, 98.80.Es, 98.62.Py}
\maketitle

\section{Introduction}

In a previous paper \cite{VFW} (henceforth VFW) we studied
spherically symmetric, dust cosmological models, which are described by the
Lema\^{i}tre-Tolman-Bondi (LTB) metric.
In particular we discussed the potential for such models to reproduce the
apparent acceleration of the Universe, and how some models evade
positivity theorems for the central deceleration parameter.
Our results were criticized in the recent paper \cite{KHCB}
by Krasinski, Hellaby, C\'{e}l\'{e}rier, and Bolejko
(henceforth KHCB).  Most of these criticisms are due to
misinterpretations and are invalid.  In this short note we clarify
some of the issues involved and respond to the criticisms.

\section{TWO KEY MISCONCEPTIONS IN KHCB}

\subsection{Misidentification of deceleration parameter}

A key issue that arises is the definition of the deceleration parameter.
Throughout VFW we defined the deceleration parameter $q(z)$ as a
function of redshift $z$ to
be what would be computed by observers from observed luminosity
distances $D_L(z)$ by assuming a spatially flat\footnote{If
one instead allows for a spatially curved geometry, the deceleration
$q(z)$ at finite $z$ can be altered, but the central deceleration
$q(0)$ at $z=0$ is unchanged, as we discussed in VFW.} Friedmann-Robertson-Walker (``FRW'' in VFW and ``FLRW'' in KHCB) cosmological model. This is discussed explicitly around our Eq. (2.10).
Specializing to $z=0$ gives the observed deceleration parameter at the
location of the observer, $q_0 = q(0)$.

This discussion applies to spherically symmetric spacetimes, with the
observer at the symmetry center.  For more general inhomogeneous spacetimes,
some prescription for angle averaging is required in order to define a
deceleration parameter.  One such definition at $z=0$ is given in
Ref.\ \cite{EF}, which in spherical symmetry coincides
with the definition of $q_0$ given above.  Slightly different
definitions, specialized to $z=0$, are given in Hirata and Seljak
\cite{HS} (henceforth HS), namely the quantities which they denote by
$q_3$ and $q_4$.  Those definitions also coincide with the above $q_0$
in spherical symmetry.

The starting point of our paper VFW was the following contradiction
that existed then in the literature.  Namely, the result of Ref.\
\cite{EF} specialized to spherical symmetry showed
that $q_0 \ge 0$ always for any smooth LTB model.
Similar results in HS related to their quantities $q_3$ and $q_4$,
specifically the argument from their Eqs.\ (18) -- (41), give the same
result when specialized to spherical symmetry.
On the other hand, the explicit LTB solutions constructed in Ref.\
\cite{INK} and elsewhere (see VFW for references) have $q_0 < 0$.
We pointed out in VFW the resolution of this apparent contradiction:
the solutions of Ref.\ \cite{INK} are not smooth at their center, and
therefore
violate the smoothness assumptions used in deriving the positivity theorems of
Refs.\ \cite{EF,HS}.

The first key error made in the comments of KHCB is a
misidentification of which definition of deceleration parameter is involved, wherein they focus on another definition given in
HS -- related to the derivative with respect to a fluid element's proper time of the expansion of the fluid -- which HS denote by $q_1$ and KHCB denote by $q_{HS}$.
One of the main criticisms in KHCB is that we were confusing different definitions of $q$ in VFW, using both $q_0$ defined above and $q_{1}$, but this is not the case as $q_{1}$ was {\it never} referred to in VFW.
In addition, when we cited HS we explicitly stated which definition of
acceleration parameter was involved, saying ``the local expansions of $\ldots$ Hirata and Seljak show that $q_0$ is constrained to be positive.''
We then proceeded to
specify the definition of deceleration parameter $q_0$ [our Eq.\ (2.16)]
and reviewed the corresponding positivity theorem.

Having focused on the incorrect deceleration parameter $q_1$,
KHCB then argued, correctly, that there is no contradiction between the
positivity theorem of HS for $q_1$ and the explicit models of
Ref.\ \cite{INK}.  This is because, although the definitions of $q_0$ and
$q_1$ coincide for smooth, spherically symmetric models at $z=0$, they
do not coincide for models which are not smooth, and in particular the
models of Ref.\ \cite{INK} have $q_1 >0$ even though $q_0 < 0$.
While KHCB are correct about this point, it is not relevant to a
discussion of VFW.

We emphasize that we used only a single definition of deceleration
parameter throughout our paper, and  KHCB's claim that we ``intermittently
used the same name$\ldots$for two different quantities'' is incorrect.
Although KHCB give two different formulae that they claim we use,
their Eqs.\ (2.12) and (2.13), in fact the first formula is just the
second formula evaluated at $z=0$.

There are further criticisms in Sec.\ VI of KHCB related to our
computation of the deceleration parameter.  After their
Eq.~(6.1) KHCB say that ``this makes it clear that the shear and
vorticity terms $\ldots$ are also evaluated at $z=0$, though this is not
clear in either paper.''  In fact this is explicitly stated in
our Ref.~\cite{EF}, where the first sentence in the
paragraph containing Eq.~(3) reads ``$\ldots$in terms of the
density, 4-velocity and velocity gradients of the cosmological fluid
evaluated at the observer's location ${\cal P}$.''  It is also
stated explicitly in VFW, where $q_0$ is defined to be the ``central
deceleration parameter'' before Eq.~(2.16),
so it is clear that everything
is evaluated where the observer lives, at $z=0$.
The fact that the gradients of the velocity are evaluated at $z=0$
also invalidates KHCB's criticism in the first paragraph of their Sec.\
VID.

\subsection{Skepticism about relevance of
  differentiability of solution}

Because there is no contradiction involving the deceleration parameter
$q_1$, KHCB believed that our explanation in terms of non-smoothness
was invalid.  They did appreciate that a contradiction still existed
for the deceleration parameter $q_0$, which they summarized in their
Sec.\ VIA.  However, they discounted
our resolution of the contradiction, and instead argued that the
formula for the deceleration parameter derived in the earlier paper
\cite{EF} was ``erroneous or erroneously interpreted in the VFW
paper.''

In order to address this skepticism, we give here a detailed and
explicit demonstration of the connection between smoothness of the
geometry and the
positivity theorems of Refs.\ \cite{EF,HS}.  Although such a
demonstration is not strictly necessary since the argument of VFW is
complete and self contained, it may nonetheless help to dispel confusion about
this issue.

A basic ingredient of those positivity theorems is a local covariant
Taylor expansion of the fluid's 4-velocity about a given point $x^\alpha$, of the form
\begin{eqnarray}
u_{\alpha'}(x') &=& g_{\alpha\alpha'}(x,x') \left[u^\alpha(x) + u^{\alpha\beta}(x)
  \sigma_{;\alpha}(x,x') \sigma_{;\beta}(x,x') \right.
\nonumber \\
\mbox{} &&
+ {1 \over 2}
u^{\alpha\beta\gamma}(x)
  \sigma_{;\alpha}(x,x') \sigma_{;\beta}(x,x') \sigma_{;\gamma}(x,x')
\nonumber \\
\mbox{} &&
  \left. + \ldots \right],
\label{eq:expansion}
\end{eqnarray}
where $u^{\alpha\beta}(x) = - \nabla^{(\alpha} u^{\beta)}(x)$ and
$u^{\alpha\beta\gamma}(x) = \nabla^{(\alpha} \nabla^\beta u^{\gamma)}(x)$.
Here $g_{\alpha\alpha'}(x,x')$ is the parallel transport bivector
and $\sigma(x,x')$ is Synge's worldfunction; see Ref. \cite{EF} for
details.  Clearly, in order for this expansion to be valid, the
coefficients $u^\alpha$, $u^{\alpha\beta}$, and $u^{\alpha\beta\gamma}$
in the expansion must exist, and in particular the symmetrized second
covariant derivative of the 4-velocity
\be
\nabla_{(\alpha} \nabla_\beta u_{\gamma)}
\
\ee
evaluated at $x^\alpha$ must exist.  We now show that this quantity
fails to exist at the center of symmetry for the LTB models with
$q_0<0$,
which explains why the positivity theorems are inapplicable to such models.

The metric of LTB models
can be written in the general form
\be
ds^2 = - dt^2 + e^{2 \alpha(t,r)} dr^2 + e^{2 \beta(t,r)} r^2
(d\theta^2 + \sin^2\theta d\varphi^2),
\label{mm}
\ee
for some functions $\alpha(t,r)$ and $\beta(t,r)$.  The fluid
4-velocity is $u_\alpha = - (dt)_\alpha$.  We assume the differential
structure on the manifold associated with the coordinates $(t,x,y,z)$,
where $(x,y,z)$ are given in terms of $(r,\theta,\varphi)$ by the
usual formulae for spherical polar coordinates \footnote{
For this differential structure the four velocity is smooth
($C^\infty$) while the metric is $C^1$ but not $C^2$.  It is possible
to find an alternative differential structure [using a coordinate
transformation of the form ${\tilde r} = r + f(t) r^2 + \ldots$
for some function $f(t)$] for which the metric is $C^2$ but not
$C^3$ and the four velocity is $C^1$ but not $C^2$.
However there is no choice of
  differential structure for which the metric is smooth, as can be
  seen by computing coordinate invariants like $\nabla_a \nabla^a R$,
  where $R$ is the Ricci scalar,
  which diverge as $r \to 0$ \cite{VFW}.}.
Then, it can be seen that both $u_\alpha$ and $u^\alpha$ are smooth
tensor fields.  However the metric $g_{\alpha\beta}$ and connection
$\nabla_\alpha$ need not be smooth, so covariant derivatives of
$u_\alpha$ need not exist.

A convenient piece of the second derivative to look at is $h^\alpha \equiv (g^{\alpha\beta} +
u^\alpha u^\beta) g^{\gamma\delta} \nabla_{(\beta} \nabla_\gamma
u_{\delta)}$.  From the metric (\ref{mm}) we then obtain
\be
h_\alpha dx^\alpha = \left[ {4 \over 3 r} ( {\dot \alpha} - {\dot
    \beta}) ( 1 + r \beta') + {\dot \alpha}' + {2 \over 3} {\dot
    \beta}' \right] dr,
\ee
where dots denote derivatives with respect to $t$ and primes denote
derivatives with respect to $r$.
We now expand the functions $\alpha$ and $\beta$ as
\begin{eqnarray}
\alpha(t,r) &=& \alpha_0(t) + \alpha_1(t) r + \alpha_2(t) r^2 + O(r^3),
\\
\beta(t,r) &=& \beta_0(t) + \beta_1(t) r + \beta_2(t) r^2 + O(r^3).
\end{eqnarray}
For all LTB models we have $\alpha_0 = \beta_0$ [see Eq.\ (2.20) of VFW],
and so we get near $r=0$ that
\be
h_\alpha dx^\alpha = {1 \over 3} \left[ 7 {\dot \alpha}_1(t) - 2 {\dot
  \beta}_1(t) + O(r) \right] dr.
\label{ss}
\ee
If the quantity in square brackets is nonzero, then $h_\alpha$ is
nonzero in the limit $r\to 0$, and points in the radial direction.
Therefore it has a direction dependent limit as $r \to 0$, i.e., the
limit does not exist.  (Correspondingly, higher order derivatives of
this quantity diverge.)

The coefficient in Eq.\ (\ref{ss}) will generically be nonvanishing for the type
of models in Ref.\ \cite{VFW} with $q_0 < 0$.  Such models were
characterized by having non-vanishing linear terms in the expansions
of the bang time function $t_0(r) = t_{00} + t_{01} r + t_{02} r^2 +
O(r^3)$ and curvature function $k(r) = k_0 + k_1 r + k_2 r^2 +
O(r^3)$.  Smoothness requires that $k_1 = t_{01} =0$.
For example, for the case $k(r)=0$ we have
\be
h_\alpha dx^\alpha = \left[ {8 t_{01}  \over 3 t^2} + O(r) \right] dr.
\ee
Therefore we see explicitly that for the models with $q_0<0$ the
second covariant derivative of the 4-velocity does not exist,
explaining why the positivity theorem does not apply.

\section{CRITICISMS RELATED TO CHOICES OF TERMINOLOGY}

Many of the criticisms in KHCB are not directed at our mathematical
results, but instead are related to the choice of terminology we
employ to describe our results.  In most cases the criticisms are
based on misinterpretations and are unfounded, but some of their
criticism has validity, since some of the terminology we used was
poorly chosen and apt to lead to confusion.  We now discuss and clarify the
relevant choices of terminology:

\begin{enumerate}

\item ``{\it Singularity:}''
In VFW we discussed in detail the nature and implications of the lack
of smoothness at the origin of the LTB models with $q_0<0$.
We called these locations singularities, in the loose sense that there were some
coordinate-invariant quantities which become singular there.
This was a poor choice of terminology, since ``singularity''
is usually used in the general relativity literature to mean geodesic
incompleteness, which does not apply here, and/or a divergence of the
Riemann tensor, which also does not apply here.
Although we did not claim that the singularity was a curvature
singularity, our terminology was confusing, as correctly pointed out
by KHCB.

KHCB also object to our describing the singularity as ``weak.''
While it is true that there are variety of different terminologies for
classifying singularities in use in general relativity,
``weak'' and ``strong'' are now in common use as referring to
whether parallel propagated components of the tidal distortion tensor diverge
or not.  The precise definitions of weak and strong are given, for
example, in Ref.\ \cite{Nolan}.  The spatial origin in the non-smooth
LTB models is weak in this sense (rather trivially, since the
components of the Riemann tensor are finite).  Therefore our
description is appropriate and the criticisms of KHCB on
this point are unfounded.

\item ``{\it Unphysical:}'' In VFW we asserted that the non-smooth LTB
models are unphysical.  KHCB disagreed by arguing
that one can smooth out the central singularity easily at $z < z_s$
for some small redshift $z_s$ without changing the predictions at
larger $z$.  We agree with this point, and in fact we mentioned it
in the concluding section of VFW.  The possibility of performing such
smoothing is well-known, see, for example, the numerical studies in
Ref.\ \cite{Yoo}.
However, the non-smooth models are still unphysical when the observer
is placed at the center, in the following sense:
Such models have considerably more fine tuning than smooth models,
since it is unlikely for an observer to live exactly on such a
point in the density distribution, and especially since to smooth out
the singularity in a manner compatible with luminosity distance observations
requires the introduction of a new, artificial, very
small lengthscale.

\item ``{\it Inverse Problem:}''
In Sec.\ II of VFW we discussed the straightforward computational
procedure for obtaining the luminosity distance $D_L(z)$
from the bang time function $t_0(r)$ of a zero-energy, LTB model.
In Sec.\ III we discussed the what we called the ``inverse problem,''
by which we meant simply the inverse process of attempting to find $t_0(r)$
from a specified $D_L(z)$.  KHCB appear to interpret the phrase
``inverse problem'' in a different and much more general sense, and
therefore their criticism on this point is unfounded.

\item ``{\it Effective Equation of State:}''
In VFW we defined the effective equation of state parameter $w_{\rm
  eff}(z)$ to be the equation of state that is obtained from the
data when assuming the usual framework of a flat FRW
cosmology.  We by no means said that it is the equation of state underlying the
LTB model itself (which is of course that of pressureless dust).
KHCB seem to be confused about this point in their
Sec.\ VIB, and appear to believe that we intended a literal
interpretation of $w_{\rm eff}(z)$.

\end{enumerate}

\section{FINDING LTB MODELS THAT YIELD A SPECIFIED LUMINOSITY DISTANCE AS A FUNCTION OF REDSHIFT}

%The inverse problem and transcritical solutions}

In Sec.\ III of VFW, we explored the problem of trying to find an
LTB model that would have the same luminosity distance $D_L(z)$ as a given FRW model with
or without dark energy.  We called this the ``inverse problem.''  We
found this problem to be complicated by a generic critical point in
the differential equations to be solved.  However we were still able
to show that ``transcritical'' LTB models could be constructed.
In Sec.\ V of their paper, KHCB criticize our discussion on a number of issues.  Their
criticism is largely unfounded.

First, KHCB appear to have misread our paper on one important point.
In Sec.\ V, they say ``VFW $\ldots$ argue that
only [dust] FRW models have $\ldots$ [a] critical point,'' and
``[VFW] say that [dust] FRW models provide examples of
transcritical solutions $\ldots$ but they fail to find other viable
[transcritical] solutions.''
These two assertions are incorrect, since in our Sec.\ IIIB we
explicitly constructed a variety of transcritical solutions that
correspond to choices of $D_L(z)$ other than dust FRW.
The origin of KHCB's misinterpretation might be the fact that
we gave the explicit formula angular diameter distance for dust FRW
models after Eq.\ (3.9).  However, this was given only as
an illustrative example.

Second, KHCB appear to misinterpret our statements about the generality
of critical points.  They argue that all light rays in physically reasonable cosmological models do
encounter critical points (local maxima of angular diameter distance as a function of redshift).
We agree with this assertion, and we never claimed otherwise in VFW.  
After our Eq. (3.13) we say ``only the special
class of transcritical solutions can extend to infinite
redshift.''  Although we did say these solutions
were ``special,'' we intended only to
mean special within the mathematical class of solutions to our
differential equations (3.4) and (3.5), and not within the space of
physically acceptable LTB models.  KHCB appear to have
misinterpreted us on this point.

Third, in VFW we argued that not every angular diameter distance
function $D_A(z)$ can be realized by zero-energy-function LTB models.
KHCB argued that generic choices of $D_A(z)$ are realizable. However,
the evidence in favor of this assertion they presented is in
the context of fully general LTB models, in which one varies both the
energy function and the bang time function, a context different from
that of VFW.
For the context studied in VFW, where one varies only the bang time
function, it is straightforward to explicitly confirm that not every
$D_A(z)$ can be realized by studying LTB models that are linear
perturbations of dust FRW and using the test outlined after Eq.\
(3.25) of VFW.  In the notation of VFW we choose
the angular diameter distance to be
\be
r_{{\rm FRW}}(z) = 3 \left[ 1 - \frac{1}{\sqrt{1+z}} \right] \left[ 1 +
    \varepsilon \delta(z) \right],
\ee
where $\varepsilon \ll 1$ is a small parameter and $\delta(z)$ is an
arbitrary function.  We choose $\delta(z)$ to vanish in a neighborhood
of $z=0$ and in a neighborhood of the critical point at $z=5/4$.
Then the linearized version of the differential equation (3.15) is
$V_1'(z) = \gamma(z) \delta'(z)$, where we have parameterized the
function $V(z)$ as
$
V(z) = 3 - 2 \sqrt{1+z} + \varepsilon
(1 + z)^{2/3}
\left[3- 2\sqrt{1+z}\right]^{-1/3}
 V_1(z)
$
and the function $\gamma(z)$ is
\be
\gamma(z) =
-\frac{2 \sqrt[3]{3-2 \sqrt{z+1}} \gamma_1(z)}{(z+1)^{2/3} \left(4 z
   \left(\sqrt{z+1}-4\right)+25 \left(\sqrt{z+1}-1\right)\right)}
\ee
with $\gamma_1(z) = -2 z^2+\left(9 \sqrt{z+1}-19\right)
z+20 \left(\sqrt{z+1}-1\right)$.
The boundary conditions for the differential equation for $V_1(z)$ are
$V_1(0) = V_1(5/4)=0$, and it follows that a transcritical solution
will be possible for a given choice of $\delta(z)$ only if
$
\int_0^{5/4} \gamma(z) \delta'(z)=0.
$
Therefore there are many choices of angular diameter distance that are
not realizable.

\section{Conclusions}

Because of recent criticism \cite{KHCB}, and to prevent future
misunderstandings, we have clarified some of the points made in VFW
\cite{VFW}.  In doing so we have shown that many of the
criticisms are unfounded.

\begin{acknowledgments}
The work of RAV was carried out at the Jet Propulsion Laboratory,
California Institute of Technology, under a contract with NASA.  EF
was supported in part by NSF grant PHY-0757735 and by NASA grant
NNX08AH27G. EF and IW are also supported in part by NSF grant PHY-0555216.  
Copyright 2009. All rights reserved. 
\end{acknowledgments}


\begin{thebibliography}{99}
\bibliographystyle{apsrev}

\bibitem{VFW} R. A. Vanderveld, \'E.\ \'E.\ Flanagan, I.\ Wasserman, Phys. Rev. D {\bf 74},
023506 (2006) [arXiv:astro-ph/0602476].

\bibitem{KHCB} A.\ Krasi´nski, C.\ Hellaby, M.N.\ C\'el\'erier and K.\
  Bolejko, arxiv:0903.4070.

\bibitem{EF} \'E.\ \'E.\ Flanagan, Phys.\ Rev.\  D {\bf 71}, 103521 (2005) [arXiv:hep-th/0503202].

\bibitem{HS} C.\ M.\ Hirata and U.\ Seljak, Phys. Rev. D {\bf 72}, 083501 (2005) [arXiv:astro-ph/0503582].

\bibitem{INK} H.\ Iguchi, T.\ Nakamura, and K.\ Nakao,
  Prog. Theor. Phys. {\bf 108}, 809 (2002).

\bibitem{Nolan} B.\ C.\ Nolan, Phys. Rev. D {\bf 60}, 024014 (1999).

\bibitem{Yoo} C. M. Yoo, T. Kai, and K. Nakao, Prog. Theor. Phys. {\bf 120}, 937 (2008) [arXiv:0807.0932v1].

\end{thebibliography}
\end{document}